# Crossing Filaments

Boris Filippov[1] and A.K. Srivastava[2]

[1]*Pushkov Institute of Terrestrial Magnetism, Ionosphere and Radio Wave Propagation, Russian Academy of Sciences, Troitsk, Moscow Region 142190, Russia*
*(e-mail:* bfilip@izmiran.troitsk.ru*)*
[2]*Aryabhatta Research Institute of Observational Sciences (ARIES), Manora Peak, Nainital-263 129, Uttarakhand, India*
*(e-mail:* aks@aries.res.in*)*

**Abstract** Solar filaments show the position of large scale polarity inversion lines and are used for the reconstruction of large-scale solar magnetic field structure on the basis of Hα synoptic charts for the periods when magnetographic measurements were not available. Sometimes crossing filaments are seen in Hα filtergrams. We analyze daily Hα filtergrams from the archive of Big Bear Solar Observatory for the period of 1999-2003 to find crossing and interacting filaments. A number of examples are presented and filament patterns are compared with photospheric magnetic field distributions. We have found that all crossing filaments reveal quadrupolar magnetic configurations of the photospheric field and presume the presence of null points in the corona.

**Keywords:** Filaments; Prominences; Magnetic fields; Reconnection

## 1. Introduction

Solar filaments follow large scale polarity inversion lines (PILs) of the photospheric magnetic field. It was well known for a long time and was tested several times manually and using also the automated techniques (Babcock and Babcock, 1955; Howard and Harvey, 1964; Smith and Ramsey, 1967; McIntosh, 1972; Snodgrass, Kress, and Wilson, 2000; Durrant, 2002; Ipson *et al.*, 2005). However, filaments are located in the corona several tens of megameters above the photospheric level. For more precise comparison between the axes of filaments (skeletons) and PILs the latter should be known at the same altitude. When detailed comparison is performed, the inversion lines are calculated at filament heights using, for example, potential magnetic field approximation (Schmidt, 1964; Filippov and Den, 2001) or by smoothing and averaging the magnetic data within a particular window (Durrant, 2002; Ipson *et al.*, 2005). Being well established, the relationship between PILs and filament locations is used for the reconstruction of large-scale solar magnetic field structure on the basis of Hα synoptic charts in periods when magnetographic measurements were not available (McIntosh, 1972; Duvall *et al.*, 1977; Makarov, 1994).

Magnetic fields that are observed in the photosphere are generated far below this level at the bottom of the convection zone by the action of dynamo mechanism (Schüssler and Schmidt, 1994). Magnetic tubes rise up to the solar surface due to magnetic buoyancy and emerge in the photosphere as sunspots and plages. Gas pressure in the higher levels of the solar atmosphere, chromosphere and corona, is low therefore magnetic fields spread to large volume forming the magnetic canopy (Giovanelli, 1980; Solanki, Livingston, and Ayres, 1994). Since major electric currents, which generate coronal magnetic fields, are located below the photosphere, nearly all coronal field lines originate from the photosphere and end in the photosphere. Therefore, an arch with both ends rooted in the photosphere is the general shape of a coronal field line. A polarity inversion line lies beneath an arcade of such coronal loops. A question arises that why filaments prefer the location above PILs? Dense plasma cannot be in stable equilibrium at the apex of a loop. For filament stability, the field lines should be curved upward, *i.e.* they should contain dips. More complex magnetic structure, like a quadrupolar configuration, or strong distortion of the field by the weight of the filament plasma are needed for dip creation (Kippenhahn and Schlüter, 1957; Hood and Anzer, 1990; Malherbe and Priest, 1983).

On the other hand, a PIL is a place where an electric current can stay in stable horizontal equilibrium. For vertical equilibrium, the diamagnetic property of the photosphere should be taken into account (Kuperus and Raadu, 1974; Van Tend and Kuperus, 1978). These circumstances have led to flux rope models of filament equilibrium (Pneuman, 1983; van Ballegooijen and Martens, 1989; Priest, Hood, and Anzer, 1989; Rust and Kumar, 1994; Aulanier and Démoulin, 1998; Chae *et al.*, 2001; Gibson and Fan, 2006). Stable equilibrium of a flux rope is maintained by photospheric dipolar magnetic field and by photospheric diamagnetism. Dense filament plasma is contained in the bottom sections of the flux rope magnetic flux tubes (Figure 1). During filament activation, plasma can spill over the upper parts of the flux tubes revealing the helical structure. The twisted filament structure is most clearly visible during the initial stage of a filament eruption.

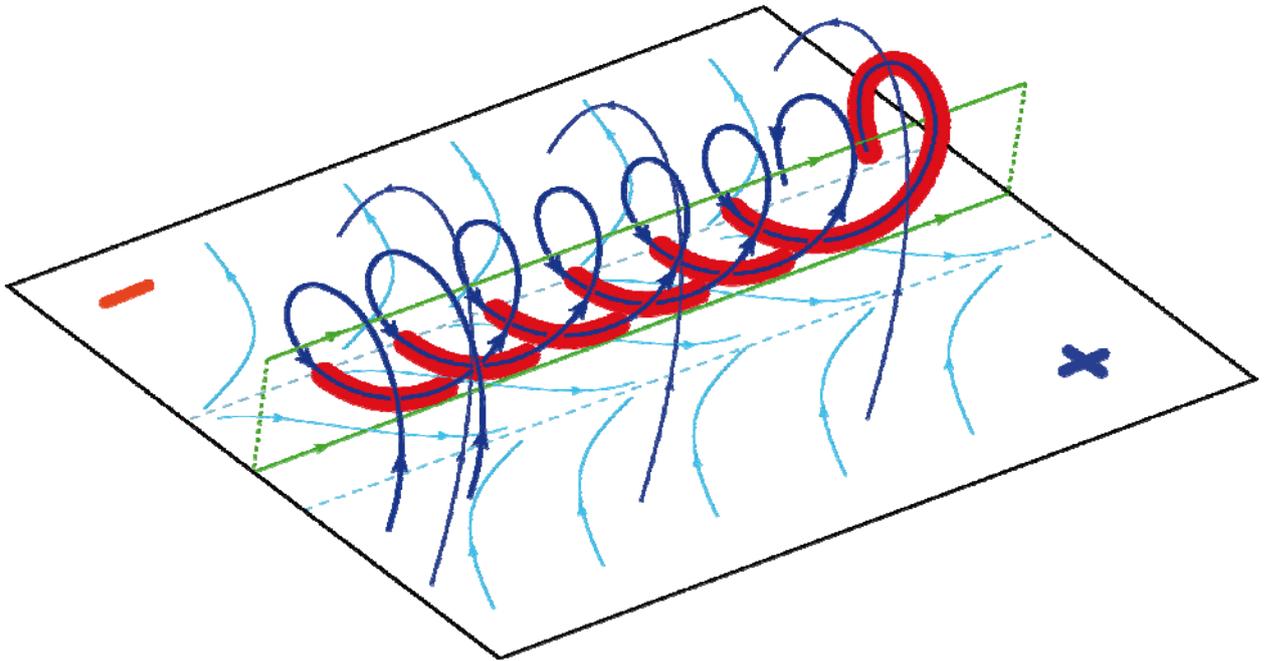

**Figure 1.** Schematic representation of a magnetic flux rope with filament plasma filling the bottom parts of the helical flux tubes. During filament activation, plasma can spill over the upper parts of the flux tubes as shown in the right-hand side of the drawing.

Filaments, being the indicators of the positions of PILs, are very helpful in the analyses of magnetic configuration in places of interest especially in regions of rather weak magnetic fields. However, sometimes filaments show pattern that is not easy to interpret. In some Hα filtergrams, crossing filaments are seen, as well as filaments combined into trident-like or three-pointed-star-like structures (Su *et al.*, 2007; Kumar, Manoharan, and Uddin, 2010). How are the magnetic polarities distributed nearby such complex structures? How ends of filaments can merge forming a joint structure? These are still the open questions.

In this paper, we analyze a series of Hα filtergrams showing complex filament arrangement. We compare filament positions with the photospheric magnetic field distribution in order to find out peculiar properties of the sites where filaments constitute unusual morphology.

## 2. Cruciform Structures

Since filaments reflect the topology of the photospheric magnetic field, filament pattern is most complicated in the epoch of maximum activity. We analyzed daily Hα filtergrams from the archive of Big Bear Solar Observatory, which includes also the data from Kanzelhoehe Solar Observatory and Yunnah Astronomical Observatory for the period of 1999-2003. Figure 2 shows examples of cruciform filament structures, when four filaments meet at a single point. Such pattern appears, if two long filaments come close together due to evolution of the photospheric magnetic field. Approaching PILs can change connectivity ("reconnect") after their contact at some point. Depending on their chiralities, pair of filaments can also reconnect or break into four independent filaments (Kumar, Manoharan, and Uddin, 2010; Filippov, 2011).

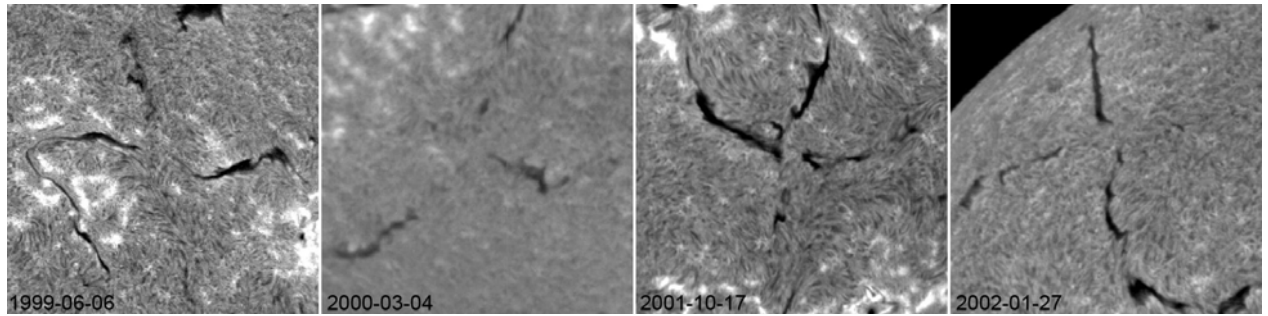

**Figure 2.** Hα filtergrams showing cruciform filament structures. The size of each frame is 450" × 450". As in all following figures, north is on top and west is on right as it is the convention for solar images. (Courtesy of Big Bear Solar Observatory).

Two intersecting PILs correspond to a quadrupole magnetic configuration (Figure 3). There is a null point near the site of intersection in the magnetic field created by photospheric sources. This field is usually assumed to be nearly potential. However, there are non-potential magnetic flux ropes stretched along PILs. The dominant component of the flux rope magnetic field is directed along a PIL (Mackey *et al*., 2010). If a pair of filaments of the same chirality comes into contact at the null point, they may exchange by their halves. During this process one segment of a filament can be connected with the two side-segments but not with the opposite segment, because the axial magnetic field components in the opposite segments are anti-parallel (Filippov, 2011). In Figure 3, the left-hand filament branch looks more connected to the upper and lower branches, while there is an apparent gap between the left-hand and right-hand filament segments.

Filaments display asymptotes of a saddle structure which can be observed around a null point (Filippov, 1995). In contrast to some chromospheric saddle structures, that may contain only a 2-D null point in the center because chromospheric fibrils show only directions of tangential magnetic fields, the null point between four orthogonal filament branches is the 3-D singular point. The normal (vertical) magnetic field component is zero, as PILs cross here. We examined SOHO/EIT images (Delaboudinière *et al*., 1995) for several days surrounding the day of observation of crossed filaments in order to find some traces of flaring activity. In all four examples shown in Figure 2, we did not find any manifestation of active processes and additional energy release near the null points. A few filament branches had erupted and disappeared. Bright arcades grew at the places of disappeared filaments but these places were rather far from null points. Kumar, Manoharan, and Uddin (2010) have also reported that flare brightenings on 20 November 2003 avoided the point of filament reconnection.

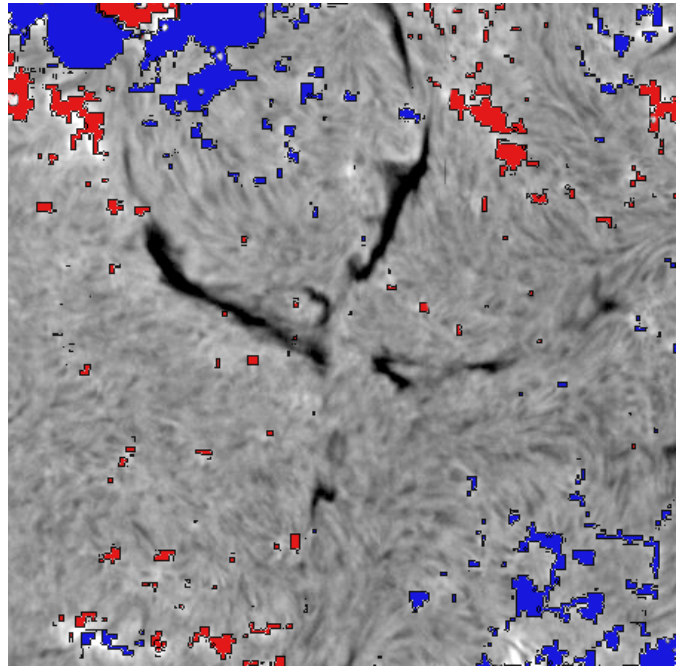

**Figure 3.** BBSO Hα filtergram on 17 October 2001 with overlaid ± 50 gauss (G) magnetic field contours from SOHO/MDI magnetogram. Red areas represent negative polarity, while blue areas represent positive polarity. (Courtesy of Big Bear Solar Observatory and SOHO/MDI consortium.)

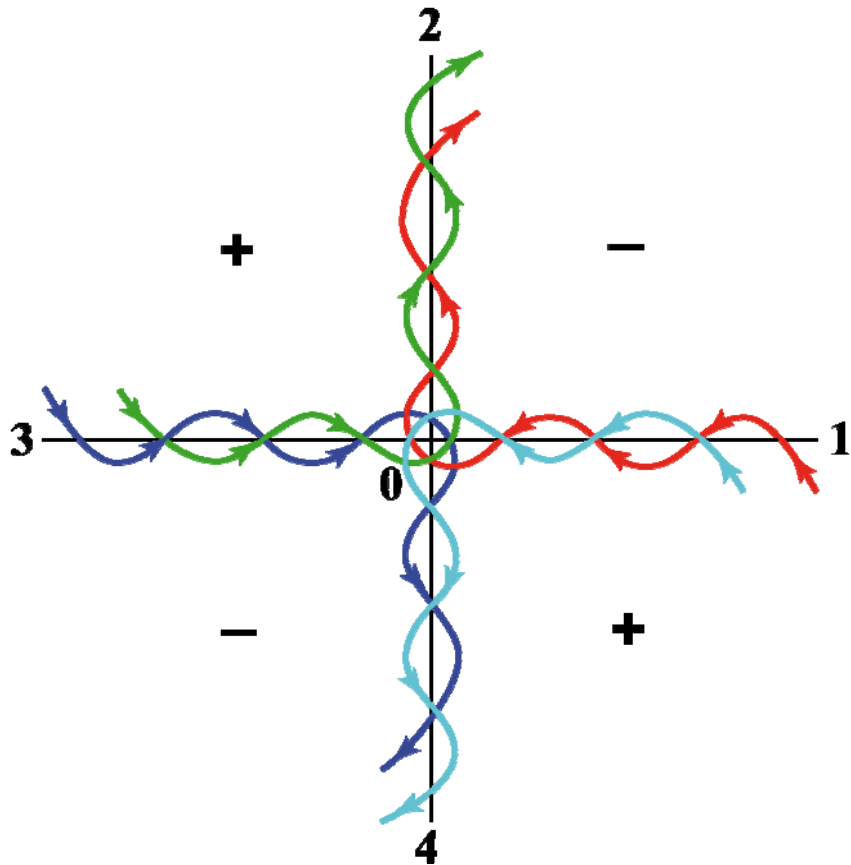

**Figure 4.** Four field lines representing four separate flux ropes within a quadrupole magnetic configuration.

In Figure 4, two intersecting PILs (or four PILs joining at a single point) are shown. In principle, it is an unstable state. Any change in the positive-negative magnetic flux balance destroys this pattern. Additional positive flux leads to the partition of PILs into two rectangular PILs with the connectivity 1-2 and 3-4. Additional negative flux leads to the partition of PILs into two rectangular PILs with the connectivity 1-4 and 2-3. However, photospheric magnetic fields change rather slowly and the degenerate state can exist many hours or even days.

Four field lines representing four separate flux ropes, which can stay in equilibrium above PILs, are shown in Figure 4 by different colors. All of them represent right-hand helices corresponding to sinistral filaments. We might say that the flux rope with the initial footpoint 1 bifurcates at the point 0 making ways to the point 2 and point 4. The same is with the three other flux ropes. There is also a possibility that some of the flux ropes (or all of them) have ending footpoints near the point 0. Since for unknown reason some flux ropes can be filled with cold and dense plasma, while others are not, the observed filament pattern can show different filament connectivity.

### 3. Three-Pointed-Star-Like Structures

More mysterious pattern is joining of three filaments at one point. If polarities are opposite on each side of every filament, then following the circle line around the central point where filaments met we do not come to the same polarity after crossing an odd number of filaments. How such structures are related to distribution of surrounding photospheric magnetic fields? It is a worth question to be explored to understand the dynamics and energetics where such complex patterns arise.

*April 1999*

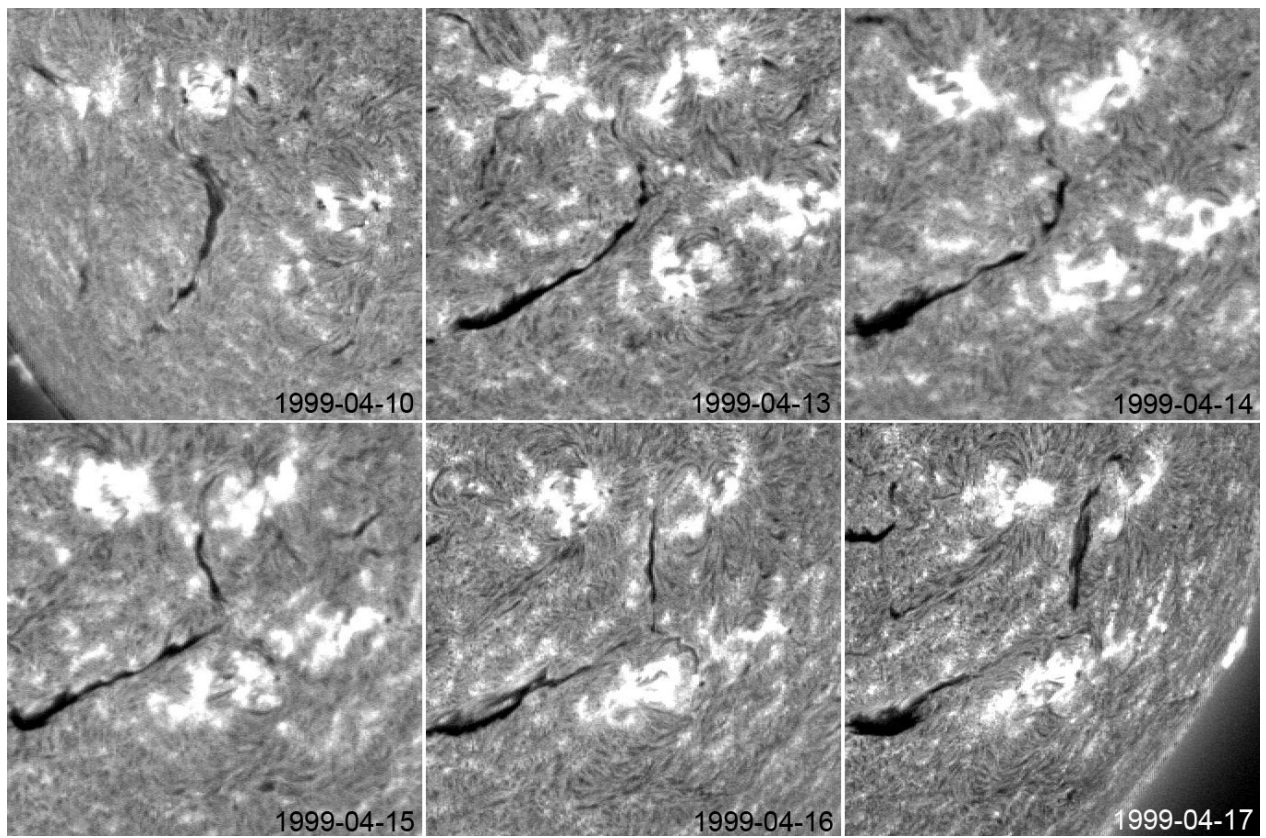

**Figure 5.** Time sequence of Hα images showing appearance of a three-pointed-star-like filament. The size of each frame is 450" × 450". (Courtesy of Big Bear Solar Observatory).

Isolated and well developed filament passed through the southern sector of the solar disk in April 1999 (Figure 5). There were no any other filaments in the vicinity of it. On 15 April, the filament axis became more curved with the right angle elbow. On the next day, a new filament branch appeared joining to the elbow. A three-pointed-star-like filament structure became easily recognized. In Figure 6, magnetic field contours ± 50 gauss (G) from SOHO/MDI magnetogram (Scherrer *et al.*, 1995) are overlaid on the Hα image. It is clearly seen that the area between the two lower filament branches is not unipolar. There are two rather compact areas of opposite polarities. Therefore, the region of the three-pointed-star-like filament structure does not have a "three-polar" magnetic configuration but a quadrupolar one. However, there is no any trace of a filament between these polarities. Comparing Figure 6 with the cartoon in Figure 4, one can conclude that the flux rope ending at point 4 does not contain filament material or is absent at all. Only flux ropes shown in Figure 4 by green and red lines correspond to visible filaments.

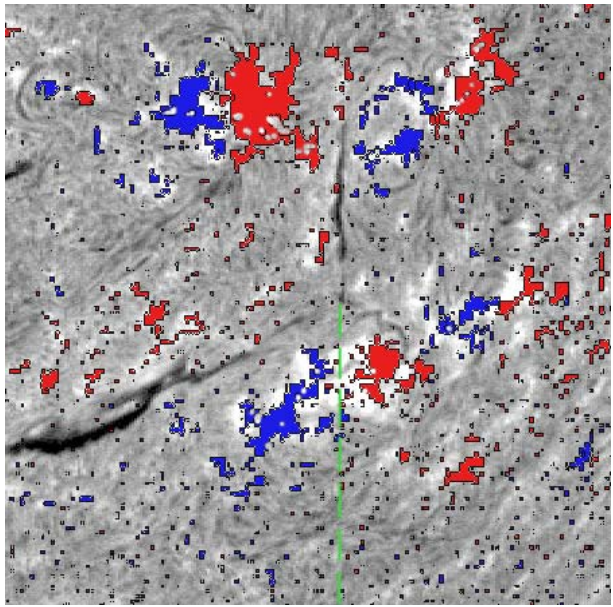

**Figure 6.** BBSO Hα filtergram on 16 April 1999 with overlaid ± 50 G magnetic field contours from SOHO/MDI magnetogram. Red areas represent negative polarity, while blue areas represent positive polarity. Green dashed line shows the approximate position of the PIL without a filament. (Courtesy of Big Bear Solar Observatory and SOHO/MDI consortium.)

*January 2001*

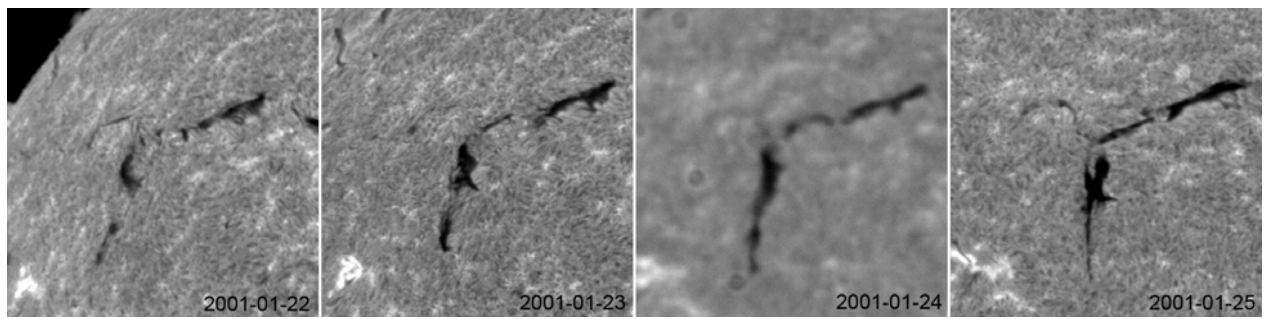

**Figure 7.** Time sequence of Hα images with a three-pointed-star-like filament. The size of each frame is 450" × 450". (Courtesy of Big Bear Solar Observatory).

Another example of filament pattern looking like a three-pointed star was observed in January 2001 in the northern hemisphere (Figure 7). All three filaments were seen rather clear in all days of the filament passage through the disk when Hα filtergrams were available. The eastern branch was fainter than the two others but it can be recognized with confidence in every image. And every day all three filaments join at one point. Comparison of filament positions with a magnetogram (Figure 8) displays that this is again a quadrupolar magnetic configuration. There are two areas of opposite polarities between the eastern and southern filament branches. In contrast to the previous case, field concentrations are small and weak in the "empty sector" but similarly a filament has not been formed along the PIL dividing them.

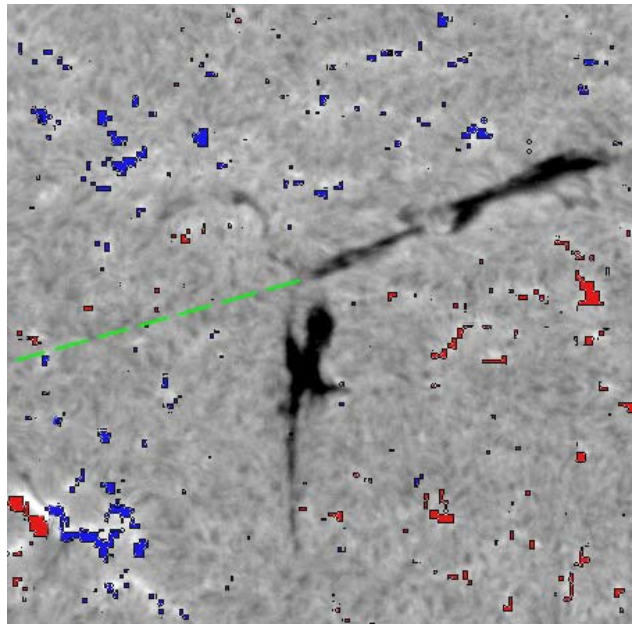

**Figure 8.** BBSO Hα filtergram on 25 January 2001 with overlaid ± 50 G magnetic field contours from SOHO/MDI magnetogram. Red areas represent negative polarity, while blue areas represent positive polarity. Green dashed line shows the approximate position of the PIL without a filament. (Courtesy of Big Bear Solar Observatory and SOHO/MDI consortium.)

### 3. Side Filaments

Sometimes a filament seems to join to a long filament from one side like a lateral road joins to a main highway. Most often such pattern is to be found at high latitudes where long chains of filaments are stretched in nearly east-west direction.

*July 1999*

Figure 9 shows a time sequence of BBSO Hα images with interlaced filaments in the northern hemisphere in July 1999. Long and rather thin filament was elongated mainly in the longitudinal direction. Wider and shorter filament joined to it from the southern side. The southern filament had high contrast and dense in the filtergrams from the day of appearance on the disk on 1 July till 6 July. It disappeared on July 7. Only faint thin traces in the southern part of the filament can be recognized. On the following days, the filament slowly reforms. Segments of the longitudinal filament close to the crossing point are more or less dense on different days. After disappearance of the southern filament on 7 July, the longitudinal filament is definitely continuous through the all its length.

Comparison of the filament pattern with a magnetogram gives at the first glance a contradictory view (Figure 10). The area to the north of the longitudinal filament is filled with elements of negative magnetic polarity. The south-western quadrant contains positive magnetic fields. In the south-eastern quadrant, there are prominent concentrations of negative polarity. The southern filament divides these two quadrants and, curving to the west, merges into the long filament. On the other hand, a clearly visible eastern segment of the longitudinal filament should divide the negative polarity on the northern side and positive polarity on the southern side. There is an area of positive polarity to the south of the filament but it is located rather far from it. The region closer to the filament contains only small elements of mixed polarities.

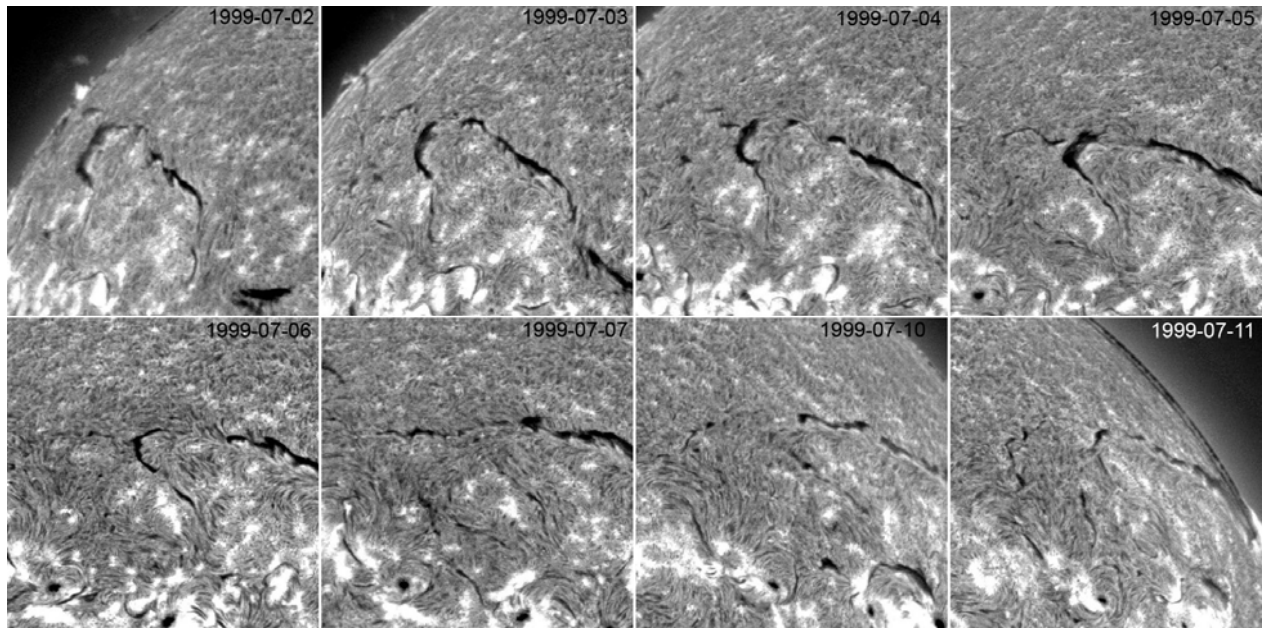

**Figure 9.** Time sequence of Hα images with interlaced filaments in July 1999. The size of each frame is 450" × 450". (Courtesy of Big Bear Solar Observatory).

To make clear the situation, we calculated the position of PILs in this region using Gaussian smoothing (Durrant, 2002; Ipson *et al.*, 2005) with different convolution kernels (Figure 10). In the top panel, the size of the convolution kernel is about 30". There is good coincidence between the position of the longitudinal (east-west) filament and the PIL. All the area on the southern side of the filament has positive polarity. A large-scaled spatial averaging corresponds to rising to a higher altitude. With the convolution kernel of about 60", we have PILs shown in the bottom panel of Figure 10. PILs have "reconnected" and, at this height, the southern filament fairly follows the PIL. Since the southern filament has a greater width than the longitudinal one, it naturally has a greater vertical size and is located at a greater height. So, it follows the higher PIL, while the longitudinal filament follows the lower PIL. There is a null point somewhere between these heights and the branch of the PIL coming out from it in the south-east direction was not occupied by a filament.

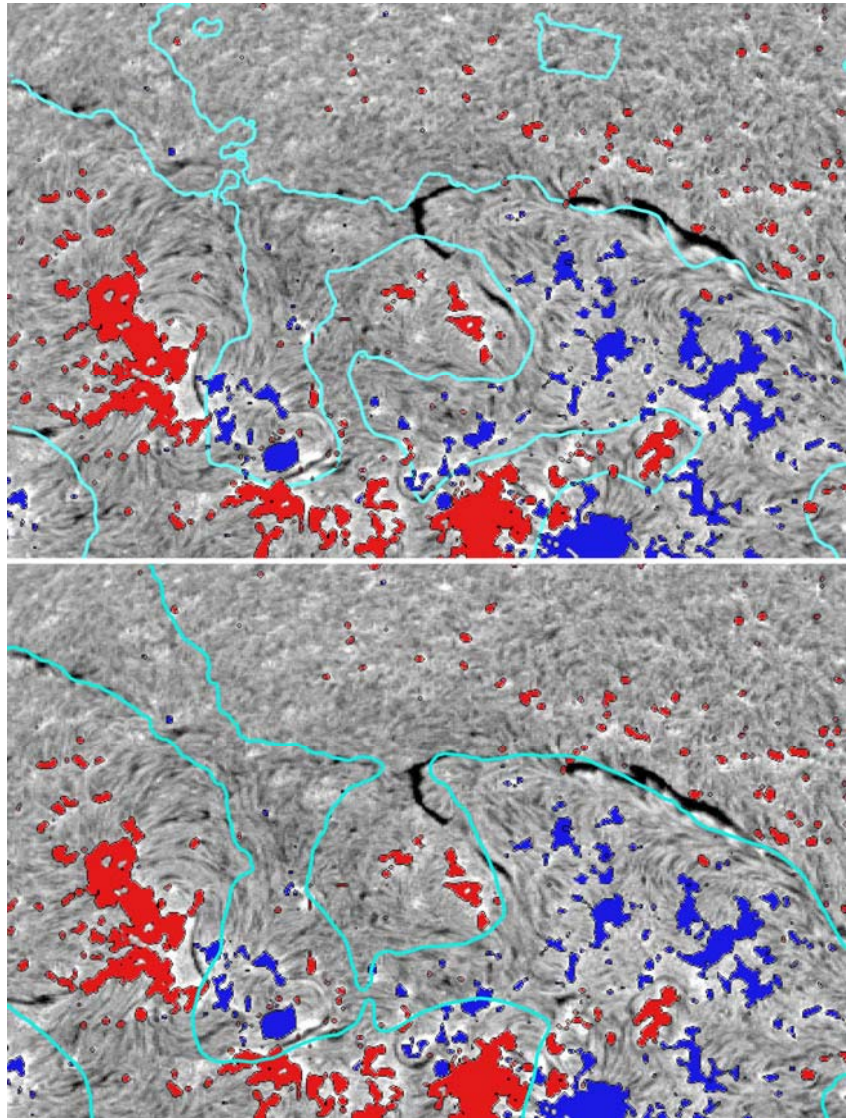

**Figure 10.** BBSO Hα filtergram on 6 July 1999 with overlaid ± 50 G magnetic field contours from SOHO/MDI magnetogram. Red areas represent negative polarity, while blue areas represent positive polarity. Cyan lines show the position of the PILs with Gaussian smoothing with 30" (top) and 60" (bottom) convolution kernels. The size of each frame is 730" × 485". (Courtesy of Big Bear Solar Observatory and SOHO/MDI consortium..)

*November 1999*

In November 1999, a well developed filament with longitudinal (east-west) orientation undergoes transformation into a curved filament with a significant portion of the body nearly perpendicular to the previous direction (Figure 11). On 9 November, the initial filament erupted. SOHO/EIT data showed that the eruption started about 14:00 UT, so the frame of 9 November in Figure 11 presents the filament during the eruption. On the next day, some part of the filament reappeared at the former place but now it had continuation to the south and even to the south-west. The filament obtained a shape of a sickle. Faint traces of a longitudinal filament to the east of the sickle are visible in a filtergram on 13 November.

An analysis of the SOHO/MDI magnetogram shows that like in the previous example, connectivity of PILs change with height. At relatively low altitude, there are two nearly parallel PILs close to each other (Figure 12). The initial longitudinal filament occupies the northern PIL. Note that the southern PIL is occupied by another filament, a small fragment of which is seen in the left-hand side of Figure 12. At higher altitude, the connectivity of PILs is different. The eastern sides of the PILs connect with each other and the western sides also do so. A shape of the western PIL is very similar to the shape of the curved filament but the position does not coincide exactly. Probably this is due to a projection effect. The latitude of the region under discussion is rather high and the height of the curved filament is presumably great. However, the averaging procedure gives us the position of the PIL as it would be seen from above. We might expect that the position of the filament coincides with the large-scale averaged PIL when the region is close to the center of the solar disk. For the region at a high latitude in the northern hemisphere a high filament should be located (in the projection against the disk) to the north of the PIL as we see in the bottom panel of Figure 12. A null point should also exist at some height between the levels of PILs presented in top and bottom panels of Figure 12. A brightening appears on 13 November not far from the position of the null point but it is related to the emergence of a new small bipolar region. The bipolar region grows significantly on the following days.

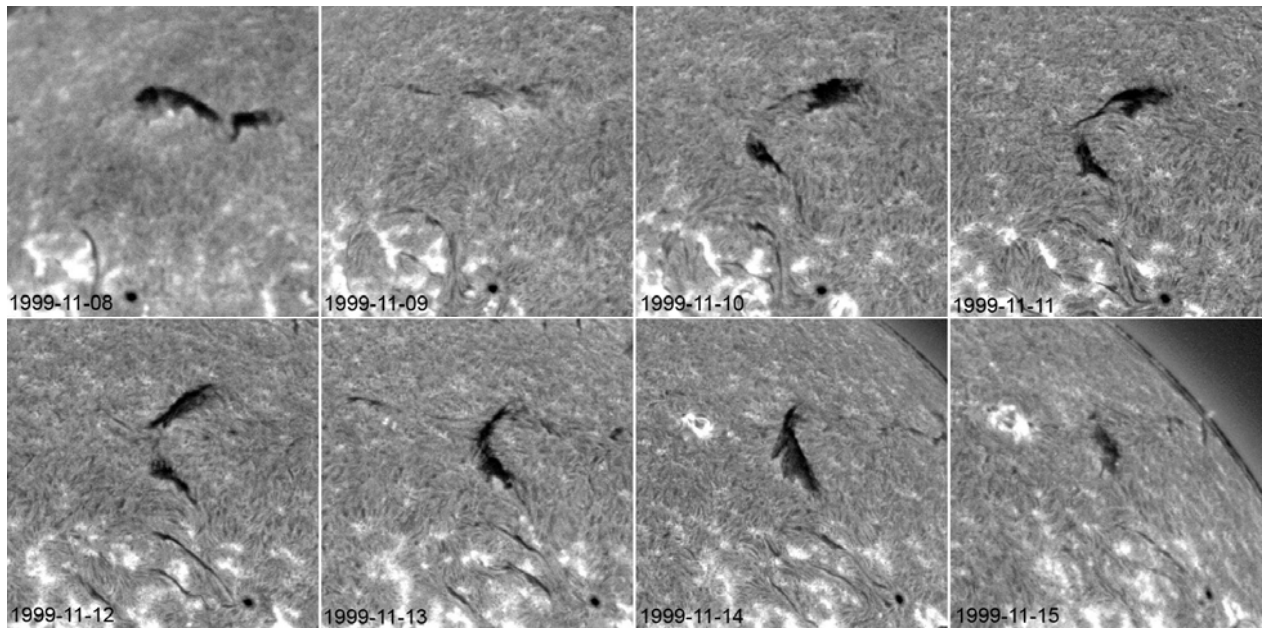

**Figure 11.** Time sequence of Hα images showing appearance of side filament in November 1999. The size of each frame is 450" × 450". (Courtesy of Big Bear Solar Observatory).

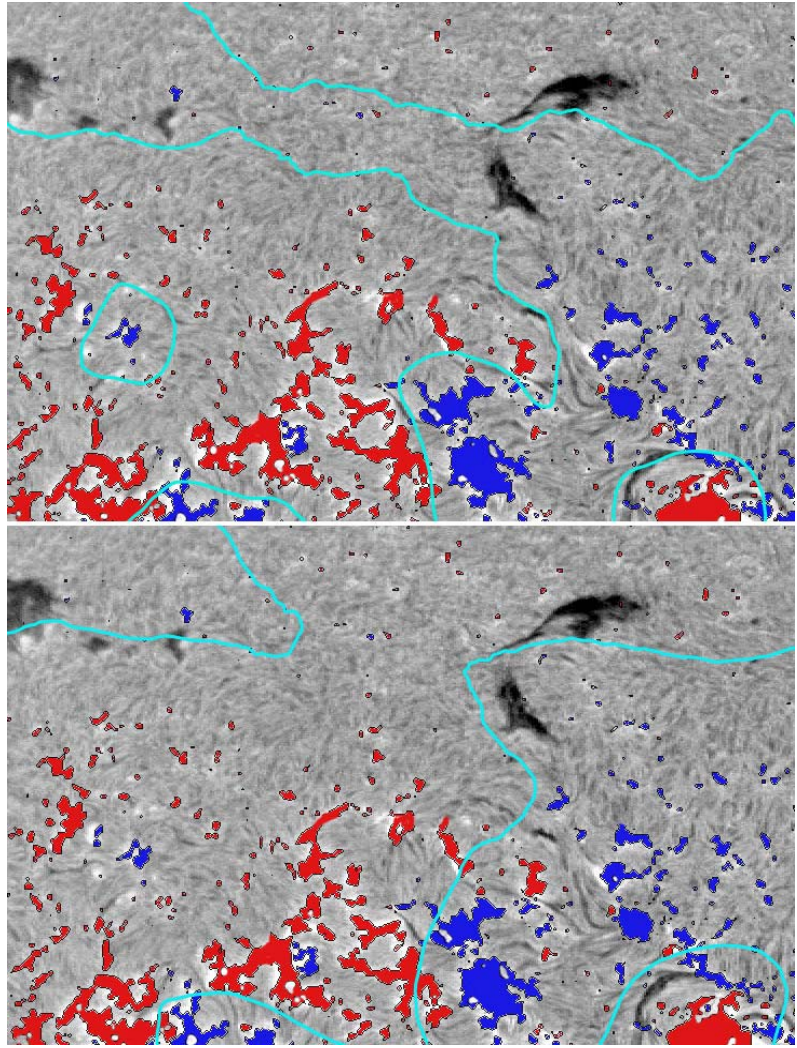

**Figure 12.** BBSO Hα filtergram on 11 November 1999 with overlaid ± 50 G magnetic field contours from SOHO/MDI magnetogram. Red areas represent negative polarity, while blue areas represent positive polarity. Cyan lines show the position of the PILs with Gaussian smoothing with 45" (top) and 90" (bottom) convolution kernels. The size of each frame is 730" × 485". (Courtesy of Big Bear Solar Observatory and SOHO/MDI consortium.)

### 4. Summary and Conclusions

We analyzed a number of examples of observations of crossing filaments in Hα filtergrams. Most transparent situation is when four filaments meet at some point. Evidently, there is a null point near the site of filament intersection. Each filament follows a PIL where the vertical magnetic field vanishes. From symmetry, the horizontal magnetic field also vanishes at the site of PILs intersection. Comparison of filtergrams with magnetograms shows that magnetic configuration is really quadrupolar at the site of filament intersection. More puzzling is a situation when three filaments join in a three-pointed-star-like structure. An analysis of magnetograms also reveals the quadrupolar magnetic configuration where one of the crossing PILs was not occupied by a filament. There are examples where a side filament merges into a long filament. Magnetic configuration is also quadrupolar in the vicinity of the filament merging. However, a null point is rather high in these cases. It is above the long filament and below the side filament. PILs have different orientation and connectivity below and above the null point. The long filament and the side filament follow PILs corresponding to their heights.

Note that we discussed earlier the magnetic configurations created by photospheric sources. This is nearly potential (current free) field. Meanwhile, there is growing conviction that filaments are related to flux ropes possessing their own non-potential magnetic field. Flux ropes can stay in stable equilibrium above photospheric PILs. Thus, filaments embedded into flux ropes are good indicators of both the flux rope presence in the corona and surrounding magnetic configurations. With the addition of the flux rope magnetic field and possibly the magnetic field of the mirror-image (by the photospheric surface) of the flux rope, the magnetic field within a filament becomes directed approximately along a PIL, or the long axis of the filament. Depending on the direction of the filament axial magnetic field relative to the surrounding magnetic configuration, filaments fall into two chirality classes: dextral and sinistral filaments (Martin, Billimoria, and Tracadas, 1994). Observations showed that two filaments of the same chirality can merge into one long filament when their ends came close to one another. Filaments of different chirality cannot merge. All filaments considered in the previous sections have the chirality dominant in the corresponding hemisphere – dextral in the northern hemisphere and sinistral in the southern hemisphere.

When four filaments are met in a null point of the photospheric magnetic field, their axial fields are antiparallel in the diagonally opposite pairs. So, the null point is not destroyed by the magnetic fields of the flux ropes. Every pair of adjacent filaments can merge because their field lines will be sewed together continuously. Reconnection of the field lines of the adjacent flux ropes can be forced by the changing of connectivity of PILs in the course of the photospheric magnetic field evolution. In studied examples, we did not find any manifestations of energy release near the null points at the places of filament intersections. However, data with better cadence are needed for detailed study of filament interactions near null points.

Physical processes near null points attract a lot of attention of solar researchers. Magnetohydrodynamic (MHD) simulations of electric current development in magnetic field configurations containing a null point have been carried out recently (Pontin and Craig, 2005; Pontin and Galsgaard, 2007; Pariat *et al.,* 2009; Masson *et al.*, 2009; Santos, Büchner, and Otto, 2010). Strong currents develop preferentially in and around the null region, along the fan plane or along the spine. Null points are expected to be locations for preferential heating by fast MHD waves (McLaughlin, Hood, and De Moortel, 2010). They are assumed by many authors as prerequisites for eruptive events and flares (Ugarte-Urra, Warren, and Winebarger, 2007; Barnes, 2007; Priest and Forbes, 2002).

Usually the position of null points in the corona is calculated using potential or force-free extrapolation of photospheric magnetograms (Wang and Wang, 1996; Régnier, Priest, and Hood, 2008). There are few observational evidences of the real existence in the corona of saddle structures corresponding to the null point surroundings (Tsuneta, 1995; Filippov, 1999). The crossing filaments are good indicators of 3-D null points in the corona. High resolution and high cadence observations of these regions can shed light on physics of current sheets and reconnection.

Study of characteristics, how filaments come into proximity, interact, change connectivity and how they are related to photospheric magnetic fields, can clarify their internal magnetic structure which is still under debates. In particular, a lot of uncertainties exist in respect of origin and evolution of the magnetic helicity of filaments, a very important parameter for estimation of geoeffectiveness of solar eruptive events.

**Acknowledgements** We thank the Big Bear Solar Observatory/New Jersey Institute of Technology and the Global High Resolution Hα Network for providing the ftp data archive. We are grateful to the SOHO MDI and EIT Consortiums for providing data. SOHO is a project of international cooperation between ESA and NASA. This work was supported in part by the Russian Foundation for Basic Research (grants 09-02-00080 and 09-02-92626) and in part by the Department of Science and Technology, Ministry of Science and Technology of India (INT/RFBR/P-38).